\documentclass[
    twocolumn,
    aps,
    prl,
    superscriptaddress,
    longbibliography,
]{revtex4-2}

\usepackage{graphicx,color,xcolor}
\usepackage[
    colorlinks=true,
    urlcolor=blue,
    linkcolor=blue,
    citecolor=blue,
    breaklinks,
]{hyperref}

\usepackage{amsmath}
\usepackage{amsfonts}
\usepackage{amssymb}

\usepackage{braket}

\usepackage[normalem]{ulem} 

\begin{document}

\title{Emergence of Fermi's Golden Rule}

\author{Tobias Micklitz}
\affiliation{Centro Brasileiro de Pesquisas F\'isicas, Rua Xavier Sigaud 150, 22290-180, Rio de Janeiro, Brazil} 

\author{Alan Morningstar}
\affiliation{Department of Physics, Princeton University, Princeton, NJ 08544, USA}

\author{Alexander Altland}
\affiliation{Institut f\"ur Theoretische Physik, Universit\"at zu K\"oln, Z\"ulpicher Str. 77, 50937 Cologne, Germany}

\author{David A. Huse}
\affiliation{Department of Physics, Princeton University, Princeton, NJ 08544, USA}
\affiliation{Institute for Advanced Study, Princeton, NJ 08540, USA}

\date{\today}

\begin{abstract}
Fermi's Golden Rule (FGR) applies in the limit where an initial quantum state is weakly coupled to a {\it continuum} of other final states overlapping its energy.  Here we investigate what happens away from this limit, where the set of final states is discrete, with a nonzero mean level spacing; this question arises in a number of recently investigated many-body systems.  For different symmetry classes, we analytically and/or numerically calculate the universal crossovers in the average decay of the initial state as the level spacing is varied, with the Golden Rule emerging in the limit of a continuum.  Among the corrections to the exponential decay of the initial state given by FGR is the appearance of the spectral form factor in the long-time regime for small but nonzero level spacing. 
\end{abstract}

\maketitle

{\it Introduction and motivation.---}Fermi's Golden Rule (FGR) describes the decay of an initial state into a continuum of final states~\cite{Dirac1927_fgr,Fermi1950_nuclear,Sakurai2017}.  To have a true continuum of final states we would need to have an infinite system, such as an excited atom emitting a photon into an infinite vacuum.  What happens if, instead, the set of final states is discrete, with some nonzero average energy-level spacing?  This question, which is addressed in the present paper, arises in multiple contexts in the study of the quantum dynamics of isolated many-body systems.

When considering the thermalization of a finite-size isolated many-body quantum
system, we ask: Does this system act as a ``bath'' for its own degrees of freedom?
As a finite-size system, it cannot be a perfect bath, since it has a discrete
spectrum.  Two limiting situations are rather clear:  If the energy-level spacing in
the putative bath is large compared to the matrix elements coupling our initial state
to these final states, then the initial state typically has no final states to
decay to that are close enough to on-shell, so the initial state typically does
not decay.  The opposite limit is where Fermi's Golden Rule does apply, because the decay
rate of the initial state is large compared to the energy-level spacing of the final
states, so the discrete spectrum serves as a good approximation to a continuum.  In
the present paper we analytically calculate the intermediate behavior between these
two limits for a single initial quantum state coupled to a chaotic quantum
dot modelled by a  random matrix of large dimension $N$, and show numerical results in full agreement with those calculations. 

The above considerations have become increasingly important recently, in an era where controlled and detailed access to mesoscopic quantum many-body systems is a reality~\cite{Monroe2021_review,Preskill2018_review,Altman2021_reveiw}. Such systems can have states that may decay into sets of other states of the system, depending on the degree to which an approximate continuum is formed for the relevant decay processes. Manifestations of
this physics, to be discussed further near the end of this paper, include: the crossover/transition at the onset of heating in periodically driven (Floquet) systems~\cite{Morningstar-Khemani2022_floquet, Seetheram-Refael2018_absence,DAlessio-Polkovnikov2013_manybody},
``avalanche'' instabilities of many-body localization (MBL) in systems with
short-range interactions~\cite{DeRoeck-Huveneers2017, Luitz-DeRoeck2017,
Thiery-DeRoeck2018}, the Fock-space localization description of the onset of many-body
quantum chaos in finite-size MBL systems with long-range
interactions~\cite{Altshuler-Levitov1997,Burin2015_long_range,Tikhonov-Mirlin2018_power_law,Altland-Micklitz2017_field_theory_mbl,Monteiro-Micklitz2021_quantum_ergodicity}, and finite-size systems with weakly broken integrability~\cite{Bulchandani-Gopalakrishnan2021_onset,Rabson-Millis2004_crossover,Znidaric2020_weak}.

In the following we will define a concrete level-dot model to use in exploring various  qualitative and universal deviations from Golden Rule behavior.  This also reveals how FGR behavior emerges from the behavior of finite systems with discrete spectra, as the limit of a continuous spectrum is approached.

{\it Model and observable.---}Consider a single level 
$|0\rangle$ weakly coupled to the 
$N\gg1$ levels
$\{|\mu\rangle\}_{\mu=1,...,N}$
  of
a fully chaotic quantum dot.
We first assume that time-reversal symmetry is broken,
and model the system by the random matrix 
Hamiltonian
\begin{align}
\hat {\cal H}
&=
\sum_{\mu,\nu=0}^N
|\mu\rangle
\begin{pmatrix}
\epsilon_0 & W\\
W^\dagger & H
\end{pmatrix}_{\mu\nu} 
\langle\nu|,
\label{eq:hamiltonian}
\end{align}
where $H$ and $W$ are 
a random hermitean $N\times N$ matrix and an $N$-component vector of independently  
distributed complex variables, respectively. 
The matrix elements have zero mean and variances
$\langle H_{kl}H^*_{mn}\rangle_H
= \frac{\lambda^2}{N}\delta_{km}\delta_{ln}$ and 
$\langle W_kW^*_l\rangle_W
= \frac{g\lambda^2}{N}\delta_{kl}$,
and we assume $\epsilon_0=0$, so the initial state resides 
at the band-center of the dot states. The spectrum of $H$ forms the Wigner semicircle~\cite{Livan2018_rmt_book}, with the density of states at its band center being $\nu = \frac{N}{\pi\lambda}$. Below we will also generalize to cases where time-reversal symmetry remains unbroken.

To study the decay of the initial state $|0\rangle$, we consider the average of the time-dependent probability to stay in that state, 
\begin{align}
\label{eq:P_res}
P(t)
&=
\langle |\langle 0 |0(t)\rangle|^2 \rangle_{H,W}.
\end{align} 
Averages $\langle... \rangle_{H,W}$ here are over the random components of $H$ and
$W$, and $|0(t)\rangle = e^{-i\hat {\cal H}t}|0\rangle$ is the time-evolved initial
state. Before turning to concrete calculations, let us formulate some qualitative anticipations. 

In the case where we assume the quantum dot is a perfect bath, Fermi's Golden Rule gives the decay $P(t)\sim e^{-\Gamma_{\rm GR}|t|}$ with the realization-averaged decay rate  
\begin{align}
\label{eq:decay_rate_FGR}
\Gamma_{\rm GR}
&=
2\pi \langle W_l W_l^* \rangle_W \nu = 2 g \lambda.
\end{align}
The above result predicts a vanishing long-time ``probability of residence''   $P_\mathrm{res}\equiv P(t\to
\infty)=0$. However, this result cannot completely describe the decay of $|0\rangle$ into a system of finite dimension, $N$: If the level-dot coupling is strong, e.g., $g = 1$, the long-time state $|0(t\to\infty)\rangle$ will be spread over the joint level-dot Hilbert space of dimension
$N+1\approx N$, i.e., there is a lower bound $P_\mathrm{res}\approx 1/N$, different from zero. For a diminished coupling, $g < 1$, the level broadens to mix with only about $\Gamma_\mathrm{GR}\nu \sim g N$ of the dot levels, so we anticipate an enhanced probability of residence $P_\mathrm{res}\sim 1/(gN)$. Further diminishing the coupling down to $g\sim \mathrm{O}(N^{-1})$
leads into a regime where the effects due to the nonzero dot level spacing become strong, and ultimately to a limit where only rare realizations have a dot level in resonant contact with the initial state. In the
process, FGR breaks down, and $P_\mathrm{res}$ approaches unity.

\begin{figure*}[t]
    \centering
    \includegraphics[width=0.8\linewidth]{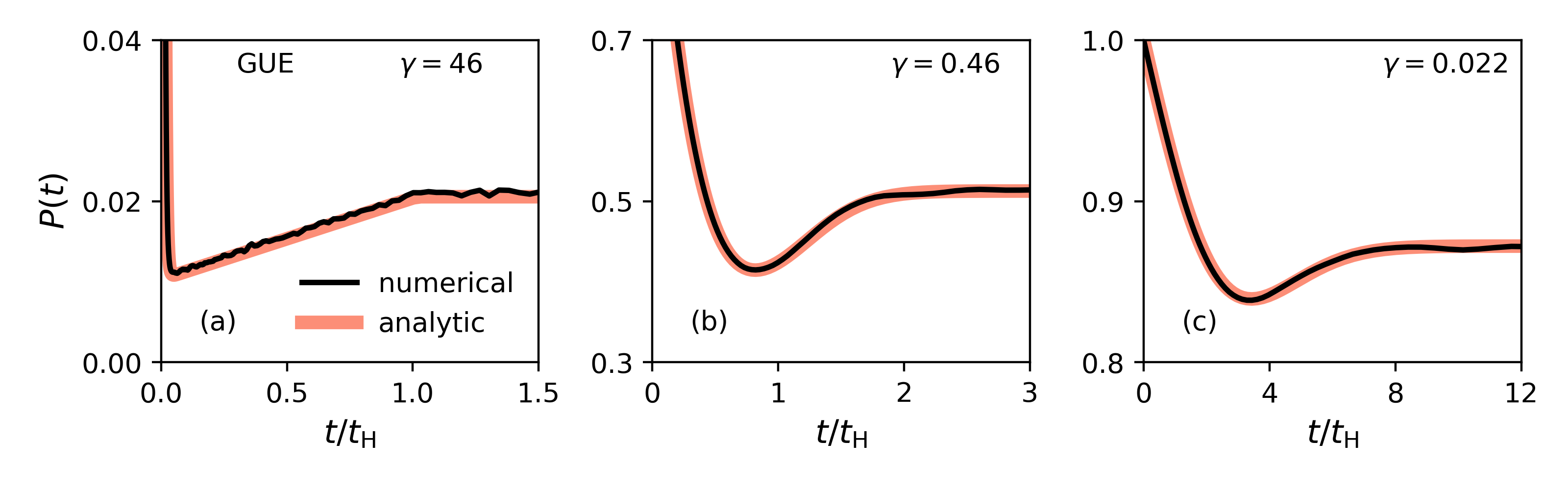}\\
    \vspace{-4mm}
    \includegraphics[width=0.4\linewidth]{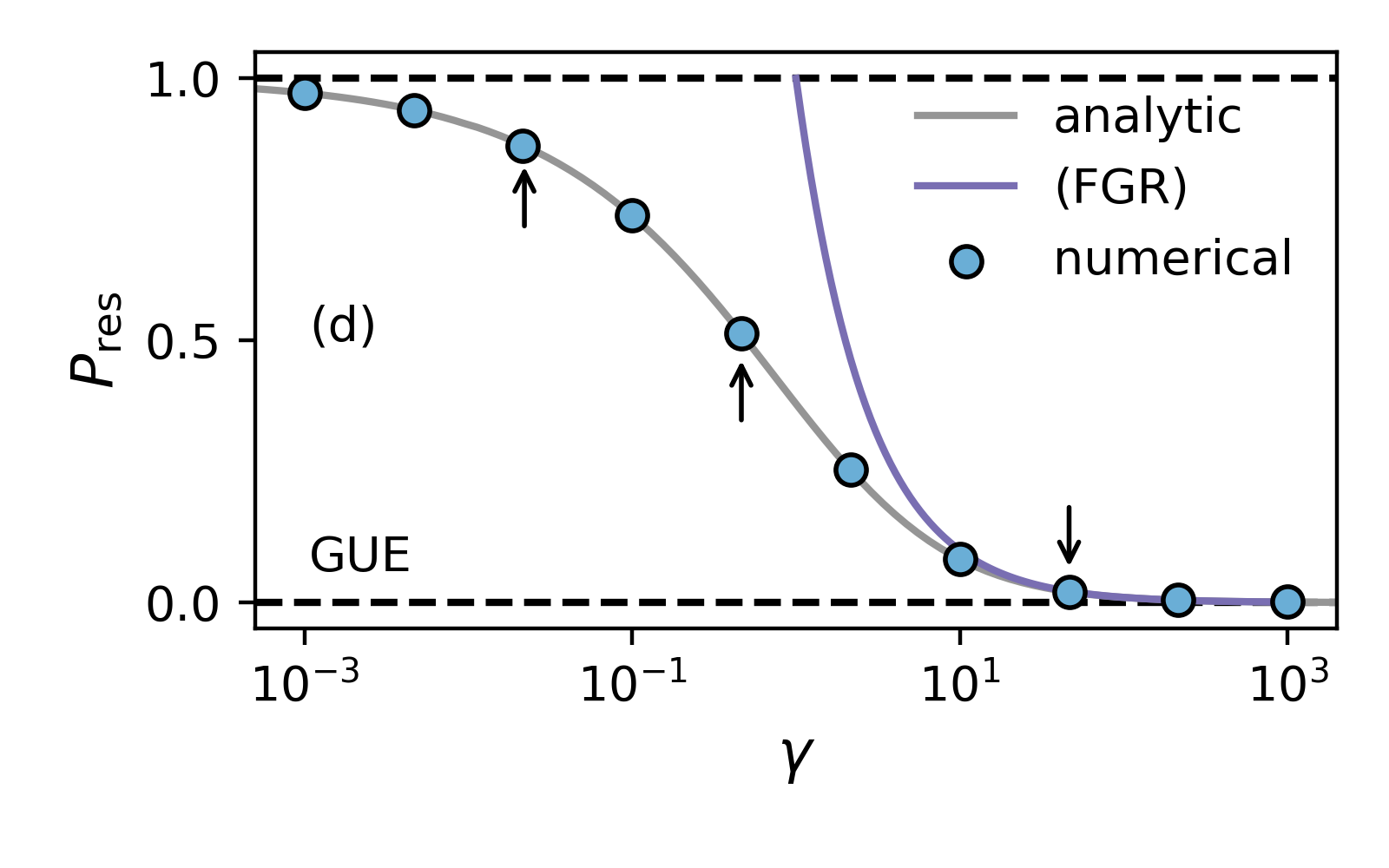}
    \includegraphics[width=0.4\linewidth]{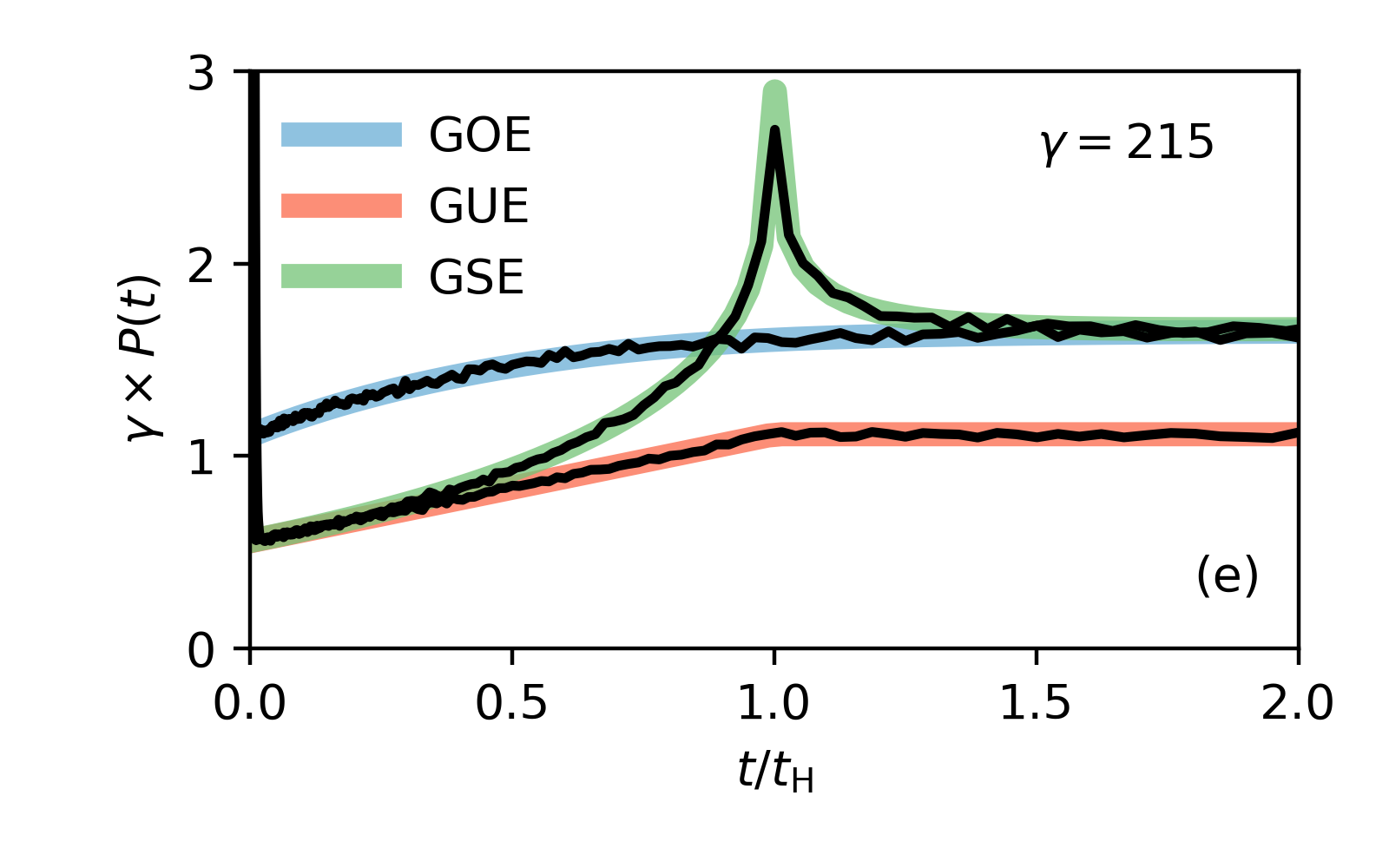}
    \vspace{-4mm}
    \caption{The probability to remain in the initial state. The total Hilbert space dimension is $N+1 = 10^3$ and numerical data are averaged over $10^4$ samples. (a)-(c) The time dependence of $P(t)$ for $\gamma\in\{46, 0.46, 0.022\}$, respectively. The black curves are data from numerical simulations of the model Eq.~(\ref{eq:hamiltonian}). The red curves are theoretical predictions obtained by evaluating the integral in Eq.~(\ref{eq:P_crossover}) numerically. (d) The $t\to\infty$ limit $P_\mathrm{res}$. Blue dots are numerical data. The gray curve is the theoretical prediction of Eq.~(\ref{eq:emergence_FGR}). The purple curve is the FGR result $1/\gamma$~\cite{finite_N_footnote}. The small black arrows point to data that corresponds to the late-time limit of the data in the three upper panels. (e) The spectral form factor arises in the time dependence of $P(t)$ for the three versions of our model derived from the three Wigner-Dyson classes (GOE, GUE, GSE). The black curves are data and the colored curves are Eq.~(\ref{eq:sff}) with parameters $a_\mathrm{X}$ and $b_\mathrm{X}$ as given in the main text. In panels (a)-(d) $H$ is GUE.}
    \label{fig:numerics}
\end{figure*}

{\it Stationary limit.---}To quantitatively describe this phenomenon, we have applied non-perturbative methods of   (effective) matrix theory; see discussion below and supplemental material. The result reads
\begin{align}
  \label{eq:emergence_FGR}
P_{\rm res}
&=
1-\gamma
-
\sqrt{\pi\gamma}\left(\frac{1}{2}-\gamma\right)
e^{\gamma}
{\rm Erfc}(\sqrt{\gamma}),
\end{align}
with $\gamma\equiv gN $ and  ${\rm Erfc}(x)=\frac{2}{\sqrt{\pi}}\int_x^\infty dx
e^{-x^2}$ the complementary error function.  The limits $N\rightarrow\infty$ and $g\rightarrow 0$ have been taken jointly such that $\gamma$ remains finite.  This formula (see the gray curve in panel (d) of Fig.~\ref{fig:numerics}) indeed predicts a crossover
from the Golden Rule estimate $P_{\rm res}=1/\gamma=1/gN$  for $\gamma>1$ (the purple curve) to a
fully decoupled level, $P_\mathrm{res}=1$, at $\gamma\to 0$. The blue dots are results obtained via numerical diagonalization for matrices of size $N+1 = 10^3$, averaged over $10^4$ samples, and are in excellent agreement with our analytic result~\cite{finite_N_footnote}.

\begin{figure}[t]
    \centering
    \includegraphics[width=8.5cm]{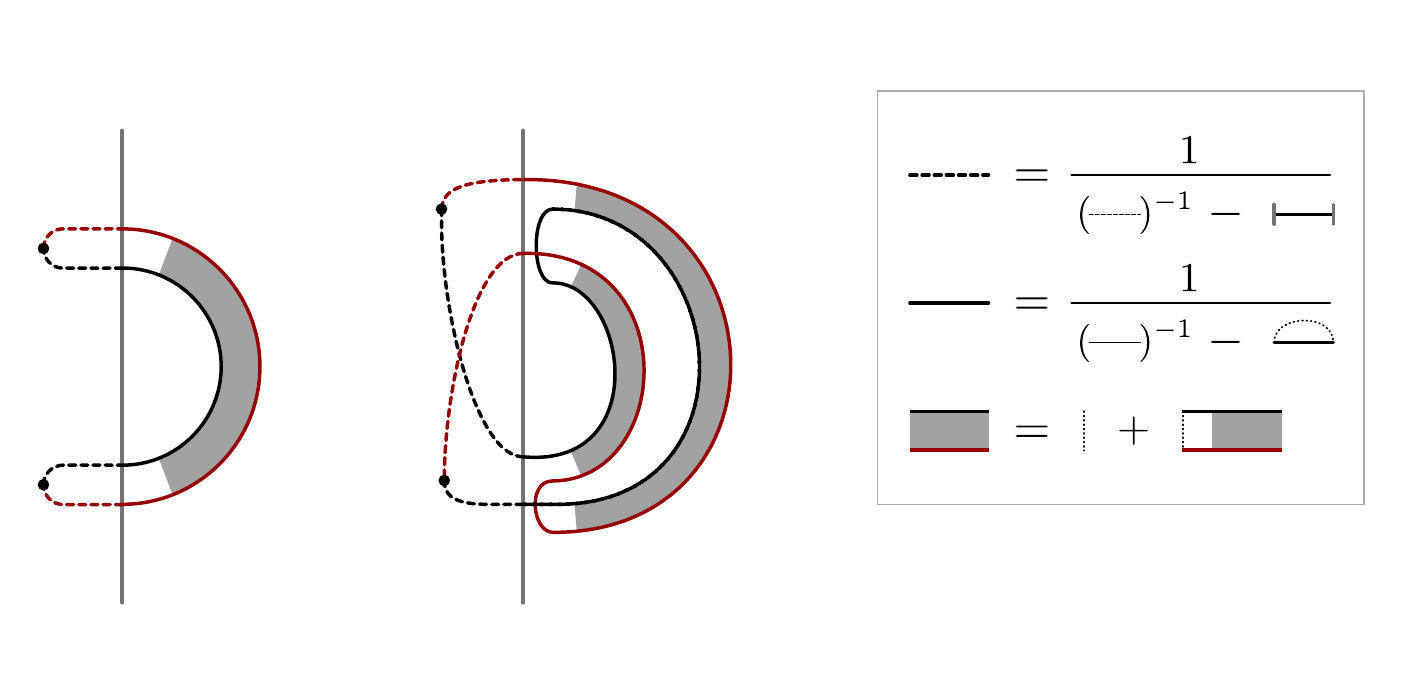}
    \vspace{-4mm}
    \caption{
    Diagrammatic representation of the short-time ($\tau< 1$) correction to the FGR exponential decay of the initial state. Left diagram (one ladder):  Classical probability that a particle, escaped into the dot, returns to the weakly coupled level. Right diagram (two ladders): Quantum interference correction to classical return probability. Dashed and solid lines here represent the retarded (black)/advanced (red) level- and dot-Green's functions 
    $G^\pm_i(\epsilon)\sim (\epsilon-\epsilon_i \pm i \Sigma_i)^{-1}$, 
     dressed by self-consistently calculated self energies $\Sigma_0\sim g\lambda$ and $\Sigma_\mu\sim \lambda$, as indicated in the box. The ladder defines the ergodic quantum dot mode $D(\omega)\sim i\lambda^2/(N\omega^+)$,  and is recursively defined in the third equation in the box. The solid vertical line represents the
     coupling between $|0\rangle$ and quantum dot states. See also the supplemental material for further details and a non-perturbative calculation addressing all times, including $\tau>1$.
    }
    \label{fig:semicclassics}
\end{figure}

{\it Dynamics.---}We next turn to the interesting question of how the above long-time limits are dynamically approached. Before turning to the quantitative computation of $P(t)$, let us discuss some estimates based on perturbation theory applied to the quantum mechanical propagator $\mathcal{P}(t)=\langle 0|0(t)\rangle=\langle 0|e^{ i\hat{\mathcal{H}}t}|0\rangle $ of the level.  In the joint limits $\gamma\to \infty$,  $g\to 0$, $\mathcal{P}(t)$ is structureless, except for incoherent FGR decay due to coupling to the
dot continuum. Substituting the broadened amplitude into Eq.~\eqref{eq:P_res}, we
obtain the zeroth-order result whose Fourier transform is $P_0(\omega)\sim 1/(\omega + 2i g\lambda)$. However, for finite $N$, coherent processes involving the propagation of the retarded
amplitude $\mathcal{P}(t)$ and its advanced complex conjugate  along identical
dot-scattering paths begin to play a role (see Fig. \ref{fig:semicclassics} for the
first two corrections of this type). While these processes are small in phase volume,
$\sim N^{-1}$, each introduces a factor in time, $t$, indicating that the effective
parameter of the perturbative analysis is $\tau \equiv t/(2\pi \nu)
\sim t\lambda/N$. Referring to the supplemental material for details, the first order process yields  the contribution $P_1(\omega)
\sim i/(gN\omega)$, where the frequency dependence $\omega^{-1}\to \Theta(t)$
reflects the ergodicity of the dot propagation on long time scales. To second order,
$P_2(\omega)\sim -\lambda/(g(N\omega)^2)$, and higher order perturbative
contributions do not exist  (in the absence of time reversal; see below).
 Fourier transformation to the time domain yields initial exponential time
dependence, $P_0(\tau)$, cut off by a constant value $P_1(\tau)\equiv P_\mathrm{off}$, followed by a linear
increase $P_2(\tau)\sim \tau$. We thus observe \emph{non-monotonic} dependence of the probability $P(\tau)$, where an initial fast exponential decay is followed by an extended period of
slow partial recovery. The resemblance of this temporal profile to that of the spectral form
factor of random matrix theory~\cite{Liu2018} suggests that the late-time increase of $P(\tau)$ will
saturate at a plateau $P(\tau)\sim (gN)^{-1}$ for $\tau>1$. However, at this stage we
are probing times $t\sim \nu$ of the order of the inverse average level spacing, outside the regime of perturbation theory.

The above discussion suggests that the discreteness of the level spacing is felt in
two different ways: saturation of $P(t)$ at times beyond the Heisenberg time,
$t_{\mathrm{H}}=2\pi \nu$, and the recovery of $P_\mathrm{res}\sim 1$ for small coupling $\gamma \lesssim 1$. To quantitatively describe these phenomena, we apply
non-perturbative methods of (effective) matrix theory (see Ref.~\cite{Efetov} and supplemental
material). The idea of this approach is to trade the complexity of an integral over
the high dimensional random matrices modeling the dot for a simpler one over a
four-dimensional (super-)matrix. The symmetry of the problem affords a further
reduction to an integral over just two ``radial coordinates'', leading to our main
analytic result
\begin{align}
\label{eq:P_crossover}
P(\tau)
=
e^{-4\gamma\tau}
&+
2\gamma^2\int_{-1}^1d\lambda_{\rm f}\int_1^\infty d\lambda_{\rm b}
\frac{x^2 e^{-2\gamma x \lambda_{\rm b}}
}{\lambda_{\rm b}-\lambda_{\rm f}}
\nonumber\\
&\quad \times
\left[
\lambda_{\rm b}I_0(z_{\rm b})
-
\mu_{\rm b}I_1(z_{\rm b})
\right]
\Theta\left(x\right),
\end{align}
where  
$x=
x(\lambda_{\rm f},\lambda_{\rm b},\tau)
=
2\tau 
-
\lambda_{\rm b}
+
\lambda_{\rm f}$, 
 $z_{\rm b}=2\gamma x\mu_{\rm b}$, with 
$\mu_{\rm b}=\sqrt{\lambda_{\rm b}^2-1}$, 
and $I_k$ are modified Bessel functions of the first kind.

For large $\gamma\gg 1$, an approximate evaluation of the integral leads to $P(\tau)
=
e^{-4\gamma\tau}
+
\frac{1}{2\gamma}
\left(1+\tau\right)
\Theta(1-\tau)
+
\frac{1}{\gamma}
\Theta(\tau-1)$, where the first two terms recover the results of our previous perturbative estimate, and the third adds the expected saturation at a plateau value. The numerical evaluation of the full integral is shown in Fig.~\ref{fig:numerics} (a)-(c) for values $\gamma=46$, $0.46$, and $0.022$ (red curves). Comparison to  numerical diagonalization of Eq.~(\ref{eq:hamiltonian}) with a total of $10^3$ levels shows excellent agreement. These curves contain the main results of our analysis: for all coupling strengths, we observe initial decay followed by a slow partial recovery of $P(t)$, which eventually terminates in a stationary plateau. Diminishing the coupling leads to a rounding of the temporal profile, and to an increase in the stationary probability with a limiting value predicted by Eq.~\eqref{eq:emergence_FGR}. 

We finally mention one more universal signature of the profile $P(t)$, namely, a
factor of two between the minimum $P_\mathrm{off}$ and the plateau value
$P_\mathrm{pl}=2P_\mathrm{off}$ in the regime $\gamma \gg 1$. This universal ratio is remarkable inasmuch as it
connects an early time semiclassical probability with a deep quantum signature
depending on the discreteness of the spectrum.  In the limits $N\gg\gamma\gg 1$, the semiclassical calculation yields $P_\mathrm{off}=1/(2\gamma)$.
The connection to $P_\mathrm{pl}$ follows from a formal  decomposition of $P(t)=\sum_{\alpha\beta} |\psi_{\alpha,0}|^2|\psi_{\beta,0}|^2 e^{i
(\epsilon_\alpha-\epsilon_\beta)t}$ in eigenfunctions. For times exceeding the
inverse level spacing, $\tau>1$, only the coherent diagonal sum $\alpha=\beta$
contributes. Noting that eigenfunctions of chaotic systems behave as Gaussian
distributed random variables, we obtain $P(t)\to
P_\mathrm{pl}=\sum_\alpha |\psi_{\alpha,0}|^4\approx 1/\gamma$, twice as high as the semiclassical value due to constructive quantum interference.

{\it Time reversal symmetry.---}While the exponential decay at short times does not depend on symmetries, later stages of the dynamics do. Distinguishing between the three cases $\mathrm{T}=(0,1,-1)\equiv(\mathrm{U},\mathrm{O},\mathrm{S})$ of: broken time reversal invariance ($\mathrm{U}$), and time reversal invariance with ($\mathrm{O}$) or without ($\mathrm{S}$) spin rotation invariance, we note that for $\mathrm{O}$, the semiclassical probability to return to the initial quantum state $|0\rangle$  doubles due to weak localization --- i.e., constructive interference between a returning path and its time reverse. No such doubling occurs for $\mathrm{S}$, because in this case the time reversed amplitude ends up in a spin reversed state that is different from the initial state. We thus expect $P_\mathrm{off}=(1,2,1)/(2\gamma)$. Turning to the asymptotic plateau value, wave functions in the presence of time reversal ($\mathrm{O}$ and $\mathrm{S}$) can be chosen real, and on this basis we expect $P_\mathrm{pl}=(2,3,3)/(2\gamma)$.  This leads to a generalization of the universal ratios as $P_\mathrm{pl} / P_\mathrm{off} = (2,3/2,3)$. To understand the temporal profile at intermediate times, we suggest the generalization of $P(t)$ at coupling $\gamma\gg 1$ to be
\begin{align}
\label{eq:sff}
P_\mathrm{X}(\tau)
&=
e^{-4\gamma\tau}
+
\frac{1}{2\gamma}
\left[
a_\mathrm{X}
+
b_\mathrm{X} K_\mathrm{X}(\tau) \right], \quad \mathrm{X}\in\{\mathrm{U}, \mathrm{O}, \mathrm{S}\},
\end{align}
i.e., a sum of incoherent decay, semiclassical return probability, and a slow partial revival described by the spectral form factor $K_X(\tau)$~\cite{Haake,Liu2018}. Assuming a normalization $K(\tau\to \infty)\to 1$, the values of the coefficients $a_\mathrm{X}=(1,2,1)$ and $b_\mathrm{X}=(1,1,2)$ follow from the requirement that $P_\mathrm{U}(\infty)=P_\mathrm{res}=\gamma^{-1}$ in the unitary case, and from the above discussion of $P_\mathrm{off}$ and $P_\mathrm{pl}/P_\mathrm{off}$. Panel (e) in Fig.~\ref{fig:numerics} shows that this hypothesis is in excellent and parameter-free agreement with our numerical analysis. In each case, a reduction of $\gamma$ will lead to a rounding of these structures as we show explicitly for the unitary case [panels (a)-(c)].  

\emph{Applications.---}Let us finally mention a few concrete contexts where we expect the above coherent generalization of FGR relaxation dynamics to be physically relevant. 
The quantum many-body phenomenon that seems to connect most directly to the results
of the present paper is the crossover/transition at the onset of heating in
periodically driven (Floquet) systems of finite
size~\cite{Morningstar-Khemani2022_floquet, Seetheram-Refael2018_absence,DAlessio-Polkovnikov2013_manybody}.  Generically, such systems are quantum
chaotic both in the non-heating regime and in the regime where they do exchange
energy with the periodic drive and thus heat up.  The basic decay process in this
case changes the system's energy by one quantum of the drive's energy, so a state of
the system at one energy decays to states at another
energy~\cite{Morningstar-Khemani2022_floquet}.

Another set of many-body systems where the discreteness of the spectrum of a putative bath plays a central role is the so-called ``avalanche'' instability of many-body localization (MBL) in systems with short-range interactions~\cite{DeRoeck-Huveneers2017, Luitz-DeRoeck2017, Thiery-DeRoeck2018}. There, a small local rare region serves as a finite bath with a discrete spectrum, and the question is whether this bath succeeds in relaxing distant spins, and thus grows in size or not~\cite{Crowley-Chandran2020_av,Potirniche-Altman2019,Goihl-Krumnow2019,Leonard-Greiner2020}.  If the coupling to the bath falls off too rapidly with the distance between the spin and the rare region, the discreteness of the finite bath's spectrum stops its ability to relax spins beyond some finite distance, so the avalanche of thermalization stops and the MBL phase can remain stable~\cite{DeRoeck-Huveneers2017, Luitz-DeRoeck2017, Thiery-DeRoeck2018}.

A third set of many-body systems where similar considerations arise is in the Fock-space localization description of the onset of many-body quantum chaos in finite-size MBL systems with long-range interactions~\cite{Altshuler-Levitov1997,Burin2015_long_range,Tikhonov-Mirlin2018_power_law,Altland-Micklitz2017_field_theory_mbl,Monteiro-Micklitz2021_quantum_ergodicity}, and relatedly, finite-size integrable systems with weak breaking of the integrability~\cite{Bulchandani-Gopalakrishnan2021_onset,Rabson-Millis2004_crossover,Znidaric2020_weak}.

{\it Summary and discussion.---}We studied the decay of a quantum state $|0\rangle$ coupled to a system with a large but, importantly, finite dimension, $N$, acting as an imperfect bath.
Assuming quantum chaos, we modelled the latter by a Gaussian-distributed random matrix of spectral range $\lambda$. We analytically calculated the probability,
$P(t)$, to remain in $|0\rangle$ at time $t$ for the case of weak coupling $g\ll1$, where the
initial state hybridizes with only $\sim\gamma = gN\ll N$ of the bath states. 
Our main observation is that the decay dynamics is generically
non-monotonic: At early times $\sim (g \lambda)^{-1}$, $P(t)$ decreases to an offset value
$P_\mathrm{off}\sim 1/(gN)$. This is then followed by a slow process of partial recovery up to a
stationary (plateau) value $P_\mathrm{pl}$, where the ratio
$P_\mathrm{pl}/P_\mathrm{off}=(2,3/2,3)$ is, for $N\gg\gamma\gg 1$, universal and depends only on the
symmetry class U, O, or S (broken time reversal, and time reversal invariant with or
without spin rotation symmetry, respectively). We established a quantitative
connection between the dynamics of recovery and the spectral form factor of quantum
chaos, thus demonstrating that the phenomenon relies on the repulsion of individual
levels in the resulting joint level-dot system. This level of sensitivity is remarkable inasmuch as the naive
Golden Rule ``smearing'' of the initial state over energy $\sim gN$ exceeds the level spacing by
far. In the opposite case where only a few levels are coupled
[$g \sim O(N^{-1})$] the decay dynamics is more complicated [Eq.~\eqref{eq:P_crossover}] and effectively described by a washed out version of the
form factor. The above phenomena are universal in that they
require only relatively mild inducers of quantum chaos. For example, we have checked numerically and analytically that randomly
distributed dot-level couplings $\{W_\mu\}$ to a bath of Poisson-distributed levels suffices to generate the above structures. 
Finally, we have identified a number of examples of genuine many-body dynamics where we expect the phase coherent generalization of Fermi's Golden Rule introduced in this paper to become physically important. However, the concrete discussion of how the fundamental physics discussed in this paper will manifest itself in such applications remains a subject of future research. 

\emph{Acknowledgments.---}D.~A.~H. was supported in part by NSF QLCI grant OMA-2120757. T.~M.~acknowledges financial support by Brazilian agencies CNPq and FAPERJ. A.~A. acknowledges partial support from the Deutsche Forschungsgemeinschaft (DFG) within the CRC network TR 183 (project grant 277101999) as part of projects A03. A.~M. was supported in part by the DARPA DRINQS program.

%

\newpage

\begin{widetext}
\section{Supplemental Material}

\subsection{Non-perturbative calculations}

\subsubsection{ Field theory}

For the non-perturbative calculation of the probability of residence 
it is convenient to start out from a more general model of two fully chaotic quantum dots 
of $N_1$ and $N_2$ levels, 
described by random hermitean matrices
from the Gaussian unitay ensemble
$\langle H_{i,kl}H^*_{i,mn}\rangle
= \frac{\lambda^2}{N_i}\delta_{km}\delta_{ln}$, 
$i=1,2$, and coupled by a random matrix $W$, with 
$g=\frac{1}{\lambda^2}{\rm tr}(WW^\dagger)$, as defined in the main text.
Employing standard supersymmetry methods 
we can calculate ensemble averaged quantities from an effective field theory with
 effective action
$S[Q_1,Q_2]
= 
S_z[Q_1]+S_z[Q_2]+S_g[Q_1,Q_2]$,
where
\begin{align}
&S_z[Q_i]
=
z_i {\rm Str}\left( \sigma_3 Q_i\right),
\qquad i=1,2,
\\
&S_g[Q_1,Q_2]
=
g {\rm Str}\left( Q_1Q_2\right),
\end{align}
with 
$z_i=-i\pi\nu_i\omega^+/2$ and 
$\pi\nu_i=N_i/\lambda$ the density of states in dot $i=1,2$.
Specifically, the probability for a particle initially prepared in dot $1$ 
to be found in the latter after a time $t$,   
\begin{align}
P_{\rm res}(t)
&=
\sum_{\nu\in {\rm dot 1}} P_{\nu\mu}(t)
=\sum_{\nu\in {\rm dot 1}} 
\langle |\langle \nu |\mu(t)\rangle|^2\rangle_{X},
\end{align} 
with average over all random matrices $X=\{H_1,H_2,W\}$, 
and 
$|\mu(t)\rangle
= e^{-i\hat {\cal H}t}|\mu\rangle$  
the time evolved initial state,
can be expressed as
\begin{align}
P_{\rm res}(t,g)
&=
\frac{F(t,g)}{F(t,0)},
\end{align} 
where
$F(t,g)
=\int_{-\infty}^\infty \frac{d\omega}{2\pi} 
F(\omega,g) e^{-i\omega  t}$ 
 the Fourier transform
 of the matrix integral 
\begin{align}
\label{app:P_res_Q1_Q2}
F(\omega,g)
&=
\int dQ_1 \int dQ_2\, e^{-S[Q_1,Q_2]} f(Q_1).
\end{align}
Here 
$f(Q)
=
{\rm Str}\left( 
Q\pi^{\rm ar}_{\rm bb} \right)
{\rm Str}\left( 
Q\pi^{\rm ra}_{\rm bb}\right)$, and 
$\pi^{\rm ss}_{\rm bb}$ projection operators onto the boson-boson sector in graded space 
and $ss'$ components in causal space, as detailed next.

\subsubsection{Matrix parametrization}

Using polar coordinates  
we parametrize $Q=UQ_0U^{-1}$, with 
\begin{align}
Q_0
&=
\begin{pmatrix}
\cos\hat\theta & i\sin\hat\theta
\\
-i\sin\hat\theta & -\cos\hat\theta
\end{pmatrix}_{\rm ra},
\qquad
 \sin\hat{\theta}
 =
 \begin{pmatrix}
 i\mu_{\rm b} & \\
 & \mu_{\rm f}
 \end{pmatrix}_{\rm bf},
\quad
 \cos\hat{\theta}
 =
 \begin{pmatrix}
 \lambda_{\rm b} & \\
 & \lambda_{\rm f}
 \end{pmatrix}_{\rm bf},
\end{align} 
where $\mu_{\rm b}\equiv\sqrt{\lambda^2_{\rm b}-1}$ 
and $\mu_{\rm f}\equiv\sqrt{1-\lambda^2_{\rm f}}$. 
Here 
$-1<\lambda_{\rm f}<1$ and $1<\lambda_{\rm b}<\infty$ are the 
compact `fermionic' and non compact `bosonic' angles~\cite{Efetov}.  
The matrix $U$ is block-diagonal in causal space and 
composed of four Grassmann variables $\eta^\pm, \bar\eta^\pm$, and two real 
variables $0\leq\phi, \chi<2\pi$, 
\begin{align}
 U
 &={\rm diag}(u_{\rm f}u_{\rm b},v)_{\rm ra}, 
 \quad
u_{\rm b}
={\rm diag}(
e^{i\phi},e^{i\chi }
)_{\rm bf},
\quad
u_{\rm f}
=e^{-2\hat \eta^+}, 
\quad
v=e^{-2i\hat \eta^-},
\quad
\hat{\eta}^\pm=
\begin{pmatrix}
0 & \bar{\eta}^\pm 
\\
-\eta^\pm & 0
\end{pmatrix}_{\rm bf}.
\end{align} 
 The integration measure in this representation reads~\cite{Efetov} 
 \begin{align}
 dQ
 &=
 \frac{1}{2^6\pi^2}\frac{1}{(\lambda_{\rm b}-\lambda_{\rm f})^2} 
 d\phi d\hat{\chi} d\lambda_{\rm b}d\lambda_{\rm f}
d\bar{\eta}^+d\eta^+d\bar{\eta}^-d\eta^-.
\end{align}

\subsubsection{Non-perturbative calculation}

On time scales larger than the 
Heisenberg time of the smaller dot 
$t\gg 2\pi\nu_1$, only a single level in the latter plays a role. 
Technically, this implies that $Q_1$-integration is dominated by 
the non compact bosonic angle (see also below), 
and we can parametrize in the action
$Q_1=\lambda_1^{\rm b}X\pi^{\rm bb}$ only by the latter, 
where $X=\left(\begin{smallmatrix} 1&-1\\1&-1\end{smallmatrix}\right)_{\rm ra}$.
Integrating over Grassmann variables of $Q_1$ we 
then arrive at the probability of residence 
\begin{align}
\label{app:P_res_gen}
P_{\rm res}(\omega,g)
&=
\pi\nu_1\int_1^\infty d\lambda_1^{\rm b}
e^{ i\pi\nu_1\omega \lambda_1^{\rm b}}
\langle e^{ - g\lambda_1^{\rm b} {\rm Str}\left( Q_2X\pi^{\rm bb}\right)}\rangle_{Q_2}, 
\end{align}
where the $Q_2$-average is with respect to the action $S_z[Q_2]$ of the isolated quantum dot 2.

\subsubsection{Integration over Grassmann variables}

Integration over Grassman 
variables of $Q_2$, the probability of residence is a sum of two contributions,
$P_{\rm res}(\omega)=P_0(\omega)+P_{\cal G}(\omega)$, 
where 
\begin{align}
P_0(\omega)
&=
\pi\nu_1
\int_1^\infty d\lambda_1^{\rm b}
e^{
\left(
i\pi\omega\nu_1
-
2g
\right)
\lambda_1^{\rm b}},
\end{align}
is a boundary term and 
 contribution from the maximal Grassmann polynomial
\begin{align}
\label{app:P-res_grassmann}
P_{\cal G}(\omega)
&=
-\nu_1g^2\partial_g
\int_0^\pi d\phi
\int_1^\infty d\lambda_1^{\rm b}
\int_{-1}^1 d\lambda_2^{\rm f}
\int_1^\infty 
d\lambda_2^{\rm b}
\frac{\lambda_1^{\rm b}}{\lambda_2^{\rm b}-\lambda_2^{\rm f}}
e^{
i\pi\nu_2\omega\left(\lambda_2^{\rm b}-\lambda_2^{\rm f}\right)
+
\lambda_1^{\rm b}
\left(i\pi\omega\nu_1
-
2g(\lambda^{\rm b}_2-\mu_2^{\rm b}\cos\phi)\right)},
\end{align}
and we notice that the boundary contribution upon Fourier transform
$P_0(t)
=
e^{-4\gamma\tau}$, where here and in the following (see also main text)
\begin{align}
\tau
&=
\frac{t}{2\pi\nu_2}, 
\qquad
\gamma
=
\frac{g\nu_2}{\nu_1}.
\end{align}
Starting out from the maximal Grassmann contribution 
Eq.~\eqref{app:P-res_grassmann} we 
integrate over angles $\phi$ to generate Bessel functions.
The Fourier transform then generates a $\delta$-function  that can be resolved by 
integration over $\lambda_1^{\rm b}$. Finally we can take derivatives and use the 
properties of Bessel functions to arrive at the result stated in the main text.

\subsubsection{Long time limit}

In the long time limit $\tau\to\infty$ 
the boundary contribution can be neglected.
We further notice that the 
leading contributions to the integral Eq.~\eqref{app:P-res_grassmann}
result from non compact angles $\lambda_1^{\rm b},\lambda_2^{\rm b}\gg1$ to simplify
$\lambda_2^{\rm b}-\lambda_2^{\rm f}\simeq \lambda_2^{\rm b}$ and 
$\lambda_2^{\rm b}-\mu_2^{\rm b}\cos\phi\simeq \lambda_2^{\rm b}\phi^2/2+1/(2\lambda_1^{\rm b})$. 
Inserting into Eq.~\eqref{app:P-res_grassmann} we can complete the trivial integral over 
$\lambda_2^{\rm f}$ and Gaussian integral over $\phi$, 
and find upon Fourier transform
\begin{align}
P_{\rm res}(t)
&=
-\frac{\nu_1g^2}{\sqrt{\pi}\nu_2}\partial_g
\frac{1}{\sqrt{g}}
\int_1^\infty d\lambda_1^{\rm b}
\int_1^\infty 
\frac{d\lambda_2^{\rm b}}{\lambda_2^{\rm b}}
\sqrt{
\frac{\lambda_1^{\rm b}}{\lambda_2^{\rm b}}  
}
e^{-\frac{g\lambda_1^{\rm b}}{\lambda_2^{\rm b}}}
\delta\left(\lambda_2^{\rm b}+\frac{\nu_1}{\nu_2}\lambda_1^{\rm b}-2\tau\right).
\end{align}
At this point it is convenient to perform the derivative $\partial_g$ and then change variables 
$u=g\lambda^{\rm b}_1/\lambda_2^{\rm b}$, $\lambda_2^{\rm b}=\lambda$ 
\begin{align}
P_{\rm res}(t)
&=
\frac{1}{\sqrt{\pi}}
\int_1^\infty d\lambda
\int_{g/\lambda}^\infty du
\frac{\sqrt{u}e^{-u}}{\gamma+u}
\left(
u+\frac{1}{2}
\right)
\delta\left(\lambda-\frac{2\tau}{1+u/\gamma}\right)
=
\frac{1}{\sqrt{\pi}}
\int_0^\infty du
\frac{\sqrt{u}e^{-u}}{\gamma+u}
\left(
u+\frac{1}{2}
\right).
\end{align}
Finally, changing variables $u=x^2$, 
we arrive at
\begin{align}
P_{\rm res}(t\to \infty)
&=
\frac{1}{\sqrt{\pi}}
 \int_{-\infty}^\infty
 dx
\frac{e^{ -x^2 }}{\gamma + x^2}
 \left(
\frac{x^2}{2}
 +
x^4 
\right)
=
\int_0^\infty dt\, e^{-\gamma t}
\frac{1+t/4}{(1+t)^{5/2}},
\end{align}
as stated in the main text.

\subsubsection{Time dependence in FGR regime}

Starting out again from 
the maximal Grassmann contribution 
Eq.~\eqref{app:P-res_grassmann}, we neglect $\nu_1\omega\lambda_1^{\rm b}$ 
and perform the Fourier transform, followed by integration over $\lambda_2^{\rm f}$,
to arrive at
\begin{align}
P^{\cal G}_{\rm res}(t,g)
&=
-\frac{\nu_1g^2}{2\pi\nu_2\tau}
\partial_g
\int_0^\pi d\phi
\int_1^\infty d\lambda_1^{\rm b}
\int_1^{2\tau+1} 
d\lambda_2^{\rm b}
\lambda_1^{\rm b}
e^{
-
2g(\lambda^{\rm b}_2-\mu_2^{\rm b}\cos\phi)\lambda_1^{\rm b}}
\Theta(\lambda_2^{\rm b}+1-2\tau)
\nonumber\\
&=
\frac{\nu_1}{4\pi g\nu_2\tau}
\int_0^\pi d\phi
\int_1^{2\tau+1} 
d\lambda_2^{\rm b}
\frac{ \Theta(\lambda_2^{\rm b}+1-2\tau)}{
(\lambda^{\rm b}_2-\mu_2^{\rm b}\cos\phi)^2}
\nonumber\\
&=
\frac{1}{4\gamma\tau}
\int_1^{2\tau+1} 
d\lambda_2^{\rm b}
\lambda_2^{\rm b}\Theta(\lambda_2^{\rm b}+1-2\tau)
\nonumber\\
&=
\frac{1}{2\gamma}
\left(
1
+\tau\right)
\Theta(1-\tau)
+
\frac{1}{\gamma} 
\Theta(\tau-1),
\end{align}
as stated in the main text.

\subsection{Perturbation theory}

We provide an estimate of the relevant contributions in the diagrammatic perturbation theory, discussed in the main text. 
For convenience of the reader 
we recall the man building blocks
introduced in the man text, 
that is, the
single level and quantum-dot Green's functions, and
the ergodic quantum dot mode, 
\begin{align}
G_0^{R/A}(\epsilon)
&=
\frac{ 1}{\epsilon \pm ig\lambda},
\qquad
G_\mu^{R/A}(\epsilon)
=
\frac{1}{\epsilon-\epsilon_\mu\pm i\lambda },
\qquad
D(\omega)
=
\frac{i\lambda^2}{\nu_2\omega}.
\end{align}
Noting then that the 
coupling of retarded/advanced single level propagators to 
the quantum dot ergodic mode 
involves the sum over 
$g=\sum_\mu |W_{\mu\nu}|^2/\lambda^2$  
on-shell quantum dot states, 
  we can estimate the three contributions discussed in the main text,
  \begin{align}
P_0(\omega)
&\sim
\int \frac{d\epsilon }{(\epsilon+\omega/2+ig\lambda)(\epsilon-\omega/2-ig\lambda)}
\sim
\frac{1}{\omega+2ig\lambda},
\\
P_1(\omega)
&\sim
\int d\epsilon 
\left[G^R_0(\epsilon)\right]^2
\left[G^A_0(\epsilon)\right]^2
\times 
g^2
\times 
D(\omega)
\sim
\frac{g^2}{(g\lambda)^3}
\times
\frac{i\lambda^2}{\nu_2\omega}
\sim
\frac{i}{gN\omega},
\\
P_2(\omega)
&\sim
\int d\epsilon 
\left[G^R_0(\epsilon)\right]^2
\left[ G^A_0(\epsilon) \right]^2
\times 
g^2
\times 
\frac{D^2(\omega)}{\lambda^2}
\sim
\frac{g^2}{(g\lambda)^3}
\times
\frac{-\lambda^2}{(\nu_2\omega)^2}
\sim
-\frac{\lambda}{g(N\omega)^2}.
\end{align}
 Finally, we notice that 
a $1/s$-expansion of the field theory approach 
reproduces the above results.

\subsection{Unit normalized spectral form factor}

To avoid ambiguity, here we state the functional forms of the unit normalized spectral form factors for Gaussian orthogonal, unitary, and symplectic ensembles~\cite{Haake,Liu2018}.
\begin{align}
    &K_\mathrm{GUE}(\tau) = \begin{cases} 
      \tau & \tau < 1, \\
      1 & \tau \ge 1,
   \end{cases}\\
    &K_\mathrm{GOE}(\tau) = \begin{cases} 
      \tau (2 - \log(2\tau + 1)) & \tau < 1, \\
      2-\tau\log\left(\frac{2\tau+1}{2\tau-1}\right) & \tau \ge 1,
   \end{cases}\\
    &K_\mathrm{GSE}(\tau) = \begin{cases} 
      \frac{\tau}{4} (2-\log|1-\tau|) & \tau < 2, \\
      1 & \tau \ge 2.
   \end{cases}
\end{align}

\end{widetext}


\begin{thebibliography}{29}
\bibitem [{fin()}]{finite_N_footnote}%
  \BibitemOpen
  \href@noop {} {}\bibinfo {note} {{Eq.~\eqref{eq:emergence_FGR} requires an
  $O(1)$ correction factor as $\gamma \to N$ for finite $N$ because of the
  nonuniform density of states of the dot.}}\BibitemShut {Stop}%
\end{thebibliography}
\end{document}